\documentclass{emulateapj}
\usepackage{graphicx}
\usepackage{amsmath}
\usepackage{verbatim}
\shorttitle{Galaxy Structure in the Rest-UV}
\shortauthors{Bond et al.}

\begin{document}

\title{THE REST-FRAME ULTRAVIOLET STRUCTURE OF $0.5 < z < 1.5$ GALAXIES}
\author{Nicholas A. Bond, Jonathan P. Gardner\altaffilmark{1}, Duilia
  F. de Mello\altaffilmark{2}, Harry I. Teplitz, Marc
  Rafelski\altaffilmark{3}, Anton M. Koekemoer\altaffilmark{4}, Dan
  Coe\altaffilmark{4}, Norman Grogin\altaffilmark{4}, Eric
  Gawiser\altaffilmark{5}, Swara Ravindranath\altaffilmark{6}, Claudia
Scarlata\altaffilmark{7}}
\altaffiltext{1}{Cosmology Laboratory (Code 665), NASA Goddard Space Flight Center, Greenbelt, MD 20771}
\altaffiltext{2}{Physics Department, The Catholic University of America, Washington, DC 20064 U.S.A.}
\altaffiltext{3}{IPAC, California Institute of Technology, Pasadena, CA 91125}
\altaffiltext{4}{Space Telescope Science Institute, 3700 San Martin Drive, Baltimore, MD 21218, U.S.A.}
\altaffiltext{5}{Department of Physics and Astronomy, Rutgers
  University, Piscataway, NJ 08854}
\altaffiltext{6}{Inter-University Centre for Astronomy and
  Astrophysics, Pune, India}
\altaffiltext{7}{Minnesota Institute for Astrophysics, School of
  Physics and Astronomy, University of Minnesota, Minneapolis, MN 55455}

\begin{abstract}

  We present the rest-frame UV wavelength dependence of the
  Petrosian-like half-light radius ($r_{50}$), and the concentration
  parameter for a sample of 198 star-forming galaxies at $0.5 < z <
  1.5$.  We find a $\sim 5$\% decrease in $r_{50}$ from $1500$~\AA\ to
  $3000$~\AA, with half-light radii at $3000$~\AA\ ranging from
  $0.6$~kpc to $6$~kpc.  We also find a decrease in concentration of
  $\sim 0.07$ ($1.9 < C_{3000} <3.9$).  The lack of a strong
  relationship between $r_{50}$ and wavelength is consistent with a
  model in which clumpy star formation is distributed over length
  scales comparable to the galaxy's rest-frame optical light. While
  the wavelength dependence of $r_{50}$ is independent of size at all
  redshifts, concentration decreases more sharply in the far-UV ($\sim
  1500$\AA) for large galaxies at $z\sim1$.  This decrease in
  concentration is caused by a flattening of the inner $\sim20$\%\ of
  the light profile in disk-like galaxies, indicating that the central
  regions have different UV colors than the rest of the galaxy.  We
  interpret this as a bulge component with older stellar populations
  and/or more dust.  The size-dependent decrease in concentration is
  less dramatic at $z\sim2$, suggesting that bulges are less dusty,
  younger, and/or less massive than the rest of the galaxy at higher
  redshifts.
\end{abstract}


\vspace{0.4in}

\section{INTRODUCTION}

Observations of galaxies at rest-frame ultraviolet wavelengths
($\lambda \sim1500$~\AA) are important for tracing the evolution of
star formation and dust obscuration.  Until recently, the study of the
structural properties of galaxies in the rest-frame ultraviolet has
focused on $z\gtrsim2$, as wavelengths $<3000$~\AA\ and redward of the
Lyman Break are easily accessible in the observed-frame optical using
the Advanced Camera for Surveys \citep[hereafter, ACS,][]{ACS} on the {\it Hubble Space
  Telescope}\footnote{Based on observations made with the NASA/ESA
  Hubble Space Telescope, obtained [from the Data Archive] at the
  Space Telescope Science Institute, which is operated by the
  Association of Universities for Research in Astronomy, Inc., under
  NASA contract NAS 5-26555. These observations are associated with
  program number HST-GO-12534.}.  Furthermore, the {\it Galaxy Evolution
  Explorer} \citep{GALEX} allows for the study of galaxy structure at
$z\lesssim0.5$ \citep[e.g.,][]{GALEX00,Heckman05,TM07}.  With the installation of the Wide Field
Camera 3 (hereafter, WFC3), including the UVIS channel, we now have the capability to
directly observe the UV emission from hot stars in galaxies at
$0.5<z<1.5$, a redshift interval that spans about one third of the
history of the Universe.

Recently published studies of the morphological properties of
$0.5<z<1.5$ galaxies are drawn largely from the Cosmic Assembly
Near-IR Deep Extraglactic Legacy Survey
\citep[CANDELS,][]{Grogin11,Koekemoer11}, which observed $\sim
0.2$~deg$^2$ of sky in the optical and near-infrared with {\it
  HST}/ACS and {\it HST}/WFC3, respectively.  In one such study,
\citet{Wuyts12} performed resolved spectral energy distribution (SED)
fitting of 323 star-forming galaxies and found that the majority of
recent star formation at $0.5 < z < 1.5$ is occurring in clumps at or
near the effective radius.  These observations are consistent with
theoretical models of gas-rich turbulent disks where clumps are
supported by infalling cold streams of gas
\citep{Bournaud07,Bournaud09}.  There are alternative models involving
mergers \citep[e.g.,][]{Robertson08}, which may be important for a
subset of galaxies at these redshifts, but fragmented structures in
sources with clear rotation curves suggest that this is not the
dominant mechanism \citep{Genzel08,FS09,Law09,Shapiro09}.

At higher redshifts, rest-frame UV imaging with ACS 
reveals that most $z > 2$ star-forming galaxies are clumpy, disturbed
and disk-like in the rest-frame UV, with only $\sim 30\%$ having light
profiles consistent with galactic spheroids
\citep[e.g.,][]{Ferguson04,EE05,Lotz06,Ravindranath06,Petty09}.  These
studies find typical half-light radii of $\sim 2$~kpc at $z \sim 2-3$
and a size evolution that scales approximately as $H^{-1}(z)$.
Although the UV wavelength dependence of galaxy structure has not been
studied at high redshift, such studies have been carried out on
well-resolved galaxies in the local universe.  \citet{TM07}
found that morphology changes occur as one observes bluer in
the UV, with galaxies becoming less concentrated, clumpier and more
asymmetric.

We can obtain a clearer picture of the young stars in galaxies at $0.5
\lesssim z \lesssim 1.5$ by studying their rest-frame UV emission at
$\lambda \sim1000-4000$~\AA.  Previous studies of star-forming
galaxies in this redshift range were performed without the aid of
observed-frame UV imaging \citep[e.g.,][]{Bruce12,Wuyts12,Wuyts13} or
with relatively shallow imaging in a single filter
\citep{VoyerPhD,Rutkowski12}.  In this paper, we use data taken as
part of a program (GO 11563: PI Teplitz) to obtain UV imaging of the
Hubble Ultra Deep Field \citep[hereafter, UVUDF][]{UVUDF} and study
intermediate-redshift galaxy structure in the F336W, F275W, and F225W
filters, complementing existing optical and near-IR measurements from
the 2012 Hubble Ultra Deep Field \citep[HUDF12,][]{HUDF12} survey.  We
use AB magnitudes throughout and assume a concordance cosmology with
$H_0=71$~km~s$^{-1}$~Mpc$^{-1}$, $\Omega_{\rm m}=0.27$, and
$\Omega_{\Lambda}=0.73$ \citep{WMAP}.  With these values, $1\arcsec =
8.0$~physical~kpc at $z=1$.

\section{DATA AND METHODOLOGY}
\label{sec:data}

The UVUDF data and the optical Hubble Ultradeep
Field \citep[UDF,][]{HUDF} are both contained within a
single deep field in the Great Observatories Origins Deep Survey
South. The new UVUDF data include imaging in three filters (F336W, F275W, and
F225W), obtained in ten visits, for a total of 30 orbits per filter.  In
addition, from the UDF, we make use of deep drizzled images taken in
the observed optical with the F435W, F606W, and F775W filters.  What
follows is a brief summary of the observation strategy, data
reduction, and catalog generation.  For much greater detail on these
procedures, see \citet[][UVUDF]{UVUDF}, \citet[][UDF]{HUDF}, and (Rafelski et al, in prep).

The first half of the UVUDF observations were taken with $2\times2$
onboard binning and without any artificial background added to the
exposures.  In the second half, in order to mitigate the effects of
charge transfer inefficiency (CTI), the observations were done without
binning and with an artificial ``post-flash''.  We use only the second
half of the obervations for the following analysis.  Although the
effects of the CTI are mitigated by the post-flash, it could
potentially still alter the apparent sizes of galaxies.  We discuss
and test this possibility in Section~\ref{subsec:CTI}.

We reduced and calibrated the individual exposures incorporating
overall bias frames, custom CTI-corrected dark frames and hot pixel
removal (to be described in Rafelski et al., in prep), flat-fielding,
and background subtraction that includes the removal of the
post-flash.  After applying a CTI correction to the raw data (using
v1.0 of the standard CTI correction
tool\footnote{http://www.stsci.edu/hst/wfc3/tools/cte\_tools}), we
combined the exposures using MultiDrizzle \citep{Koekemoer02}, with a
pixfrac of $0.8$ and a square kernel, to produce final drizzled images
with a pixel scale of $30$~mas.  Details of the image combination are
the same as those described in \citet{Koekemoer11}.  In order to
ensure accurate alignment of the UV imaging with the UDF
\citep[][]{HUDF}, the WFC3 exposures were individually
aligned to the UDF B-band catalog.

We construct a source catalog (Rafelski et al, in prep),
following a procedure similar to the Ultradeep Field catalog
\citep{Coe06}.  All objects are detected using
SExtractor\citep{SExtractor} on an image that is the
weighted sum of the F435W, F606W, F775W, F850LP ACS images and F105W,
F125W, F140W, and F160W WFC3/IR images.  This detection image is used
to derive aperture corrections, which are then applied to isophotal
magnitudes measured in the UV images.
The effective depth; that is, the limiting magnitude at which sources have
a $>50$\%\ detection efficiency, is $m_{\rm AB}=27.7$, $27.7$, $28.2$ for
F225W, F275W, and F336W, respectively.

\begin{figure}[t]
\plotone{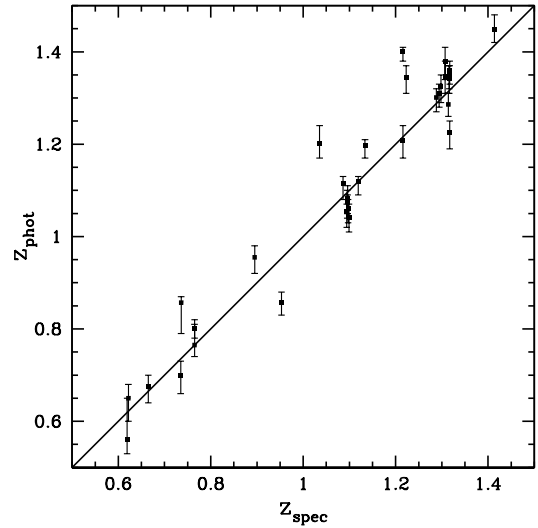}
\caption{The photometric redshift plotted against the
  spectroscopic redshift for $33$ galaxies in the UVUDF.  The photometric redshifts have a scatter in
$\Delta z/(1+z)$ of $\sigma_z=0.030$, after excluding one 3-$\sigma$
outlier, and the mean has a bias of $\Delta z/(1+z)=0.008 \pm 0.006$.
\label{fig:z_vs_z}}
\end{figure}

For all individual galaxies, we determined redshifts using a Bayesian
photo-z (BPZ) algorithm \citep{Benitez00,Benitez04,Coe06}, but the
UVUDF catalog also includes spectroscopic redshifts, compiled by
\citet{Rafelski09} from a range of sources
\citep[][]{Fevre04,Szokoly04,Vanzella05,Vanzella06,Vanzella08,Popesso09}.
We also include new measurements from \citet{Balestra10} and
\citet{GMASS}, giving a total of $33$ $0.5<z<1.5$ galaxies with
spectroscopic redshifts.  We demonstrate the accuracy of the
photometric redshifts in Figure~\ref{fig:z_vs_z} by comparing them to
spectroscopic redshifts for these galaxies.  The photometric redshifts
have a scatter in $\Delta z/(1+z)$ of $\sigma_z=0.030$ over the range,
$0.5<z<1.5$.  In addition, there is a bias of $(z_{\rm phot}-z_{\rm
  spec})/(1+z)=0.008 \pm 0.006$, which we find to be insignificant at
$<2\sigma$ using 1000 bootstrap simulations.  We also used 125 G141
grism redshifts from the 3D-HST project \citep{3DHST}, which agree
very well with the other spectroscopic redshifts, $\sigma_z=0.0056$.
For all samples here, we assume a galaxy is at its spectroscopic
redshift, if one is available.  If not, it is given either a grism
redshift or, if no other measurement is available, a photometric
redshift.

\subsection{Sample Selection}
\label{subsec:sampleselect}

\begin{figure}[t]
\plotone{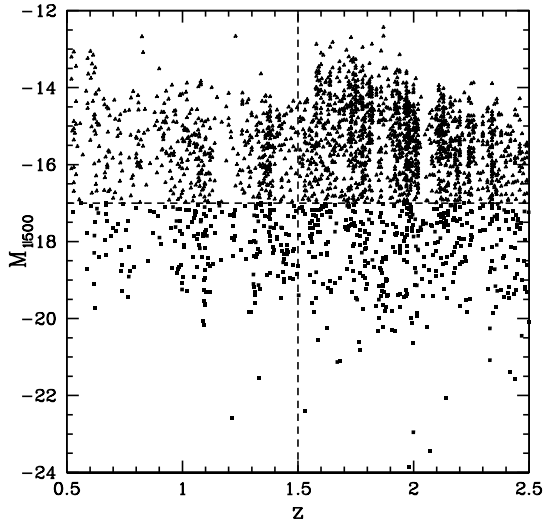}
\caption{The distribution of absolute magnitude at $1500$~\AA\ as a
  function of redshift (see Section~\ref{sec:data}) for all sources in the UVUDF catalog.  We
  separate our sample into two redshifts bins (vertical dashed line) and
  implement a cut at $M_{1500}<-17$ (horizontal dashed line) to ensure a sample that is
  luminosity-limited in the UV and for which simple morphological
  diagnostics can be measured (average S/N per pixel~$\gtrsim2$).  Note
  that, at $1.5<z<2.5$, $1500$~\AA\ is sampled primarily by the deeper
  UDF optical imaging.
\label{fig:MagSelect}}
\end{figure}

We analyze two UV-luminosity-limited samples, one at $0.5 < z < 1.5$
and another at $1.5<z<2.5$.  We select each according to the
rest-frame $1500$~\AA\ absolute magnitude, $M_{1500}$, which is
estimated from a linear interpolation between the two nearest UV or
optical magnitudes.  In Figure~\ref{fig:MagSelect}, we show the
distribution of $M_{1500}$ as a function of redshift for galaxies in
the UVUDF area.  We implement a $M_{1500}<-17$ cut to ensure
completeness at all redshifts and sufficient signal-to-noise (S/N)
that concentration measurements can be made in most bandpasses (S/N
per pixel~$>2$, see Section~\ref{subsec:morphmeasure} for more
detail on the signal-to-noise requirements).  Our source catalog is
$100$\% complete for all galaxies brighter than this magnitude cut.
The final luminosity-limited samples contain 198 galaxies at $0.5 < z
< 1.5$ and 400 galaxies at $1.5 < z < 2.5$.

\begin{figure}[t]
\plotone{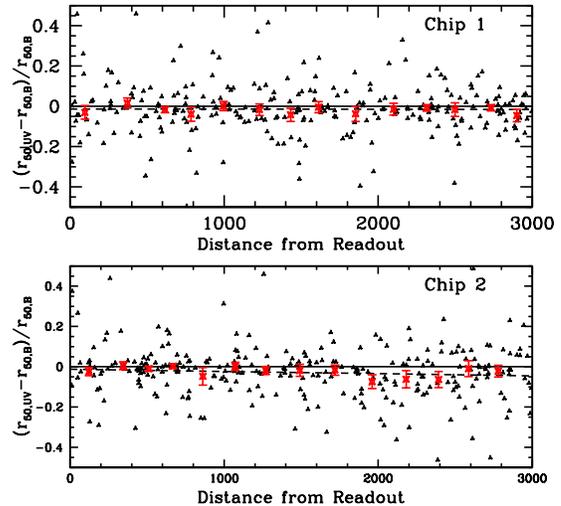}
\caption{
Fractional difference in $r_{50}$ as a function of distance from the
CCD readout for UVIS Chip 1 (top) and UVIS Chip 2 (bottom).   Black
triangles indicate individual galaxies, while red square points
indicate the median differential radius in 10-object bins. Each panel includes a
linear least squares fit (dashed line) to the data,
both of which are consistent with the zero line (solid line) at $<1.5\sigma$. If the CTI
were having a significant impact on the $r_{50}$ measurements, we
would expect them to systematically decrease approaching the readout.  
No such effect is observed.
\label{fig:SizeReadout}}
\end{figure}

\subsection{Structural Diagnostics}
\label{subsec:morphmeasure}

We measure galaxy sizes with a Petrosian-like radius \citep{Petrosian};
specifically, the radial distance at which the local surface
brightness is half of the internal surface brightness, $r_{50}\equiv r(\eta=0.5)$,
and $\eta(r)\equiv I(r)/\left<I(<r)\right>$.  This quantity is
approximately equal to the half-light radius.  We define the
concentration following \citet{Conc1} and \citet{CAS},
\begin{equation}
C=5\mathrm {log}\left[\frac{r_{80\%}}{r_{20\%}}\right],
\end{equation}
where $(r_{80\%},r_{20\%})$ are the radii at which the integrated
light profile is at 80\% and 20\% of the light within $r(\eta=0.2)$.  Both
quantities are defined in terms of a surface brightness ratio, so are
insensitive to the depth of the imaging.  

In order to access the rest-frame ultraviolet in $0.5 < z < 1.5$
galaxies, we use filters in the observed-frame near-UV and optical,
including F225W, F275W, and F336W from the UVUDF and F435W, F606W, and
F775W from the UDF.  All observed-frame UV structural measurements use
the peak flux in the optical/near-IR detection image as the center of
the galaxy (see Section~\ref{sec:data}).  We checked that our results
were insensitive to the filter used to center the galaxy, alternately using the F336W
and B-band centroids as a reference position for the structural
diagnostics.  No qualitative change in our results is observed.  Note
that in order to achieve maximum spatial resolution, we do not match
the point spread functions (PSFs) between the ACS and UVIS images.
The PSF in the optical ACS images is $\sim 10$\% larger than that in
the UVIS images, but most of the galaxies are well resolved and we
show in Section~\ref{sec:radresult} that such effects are not large
enough to bias our results significantly.

In order to test the dependence of our structural diagnostics on
S/N, we used object-by-object Monte Carlo simulations.  To
do this, we first extract the galaxies in our sample from the
appropriate ACS image (nearest to $\lambda_r=3000$~\AA) using
SExtractor (DETECT\_MINAREA~$=5$ and DETECT\_THRESH~$=1.65$).  The ACS
images are much deeper than the corresponding WFC3 observations, and
are therefore a good approximation of the galaxy's ``true'' light
distribution.  We then simulated a noise-free WFC3 UVIS observation by
normalizing this cutout so that the object's aperture-corrected
magnitude is the same as that in the real UV image.  Noise was then
added using actual realizations from blank regions in the epoch 3 WFC3
image.  We find that there are systematic decreases in both $r_e$ and
$C$ at very low S/N, but these decreases are $<1$\%
when the average S/N pixel is $\gtrsim 2$ within a
1-pixel annulus centered at $r(\eta=0.5)$ and $r_{80\%}$,
respectively.  As such, we restrict our analyses in the UV to objects
brighter than this limit.

\begin{figure}[t]
\epsscale{1.3}
\plotone{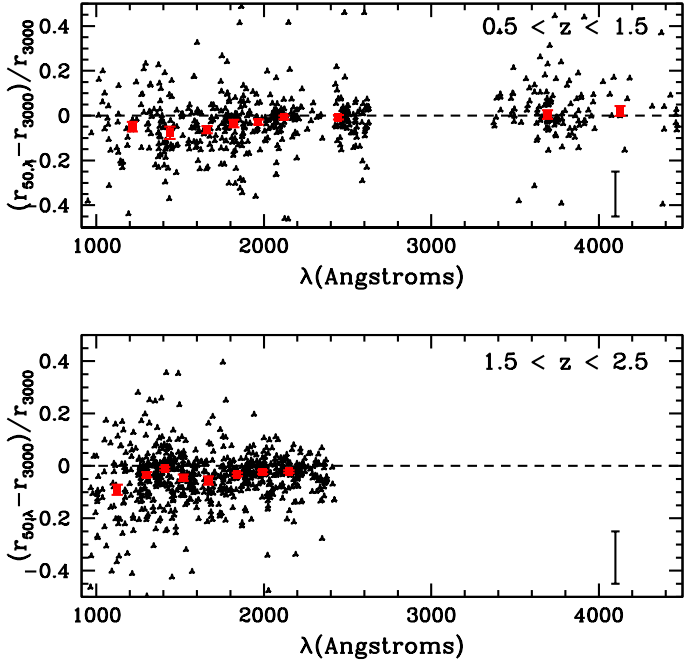}
\caption{The difference in $r_{50}$ as a function of
  rest-frame wavelength for a sample of $198$ $0.5 < z < 1.5$ galaxies
  with $M_{1500}<-17$ (top) and a sample of $400$ $1.5 < z < 2.5$
  galaxies with the same magnitude cutoff (bottom).  Differences are
  taken relative to the filter nearest to rest-frame $3000$~\AA.
  Black triangles indicate individual measurements, while red square
  points indicate the median differential radius in 10-object bins.
  Error bars for the medians are derived from 1000 bootstrap
  simulations and a typical error bar for the individual measurements
  is given in the lower right corner. The median $r_{50}$
  exhibits a decrease at the bluest wavelengths, but remains constant
  to $<15$\% over the range, $1200 < \lambda_r <
  4000$~\AA.\label{fig:Rdiff}}.
\end{figure}

\subsection{Charge transfer inefficiency}
\label{subsec:CTI}

One of the downsides to space-based CCD imaging is that the cumulative
damage from cosmic radiation, in the form of charged particles, can
create ``charge traps'' in the detector.  When charge is transferred
across the CCD in the readout phase, it can get caught in these traps,
leading to a systematic loss of source flux in the processed image, as
well as the creation of trails as the charge is eventually released
later in the readout phase.  This charge transfer inefficiency
is especially problematic in UV images, where the sky background is
low and the majority of the trapped charges will be from individually
detected sources.  In galaxies detected at moderate-to-high
S/N, the CTI effects can be corrected, but sources
at the faint end can be lost entirely due to these effects.  The
impact of CTI on galaxy shape measurements has been studied in the
context of weak lensing \citep{CTI}, and it was found that galaxies
far from the CCD readout can have their ellipticities altered by CTI effects.

For a detailed discussion of the impact of CTI on the UVUDF images and
catalog, see \citet{UVUDF}.  In Figure~\ref{fig:SizeReadout}, we plot
the fractional difference in $r_{50}$ as a function of distance from
the CCD readout in the UVUDF observations.  Each panel includes a
linear least squares fit to the data (dashed line), both with
slopes and y-intercepts that are consistent with zero at $<1.5\sigma$.
We therefore infer that there is no systematic change in $r_{50}$ as a
function of distance from the readout, as one might expect if CTI
effects were significant.  Note that both the post-flash and the
pixel-by-pixel CTI correction mitigate the effects of the CTI in these
observations.

\begin{figure}[t]
\epsscale{1.3}
\plotone{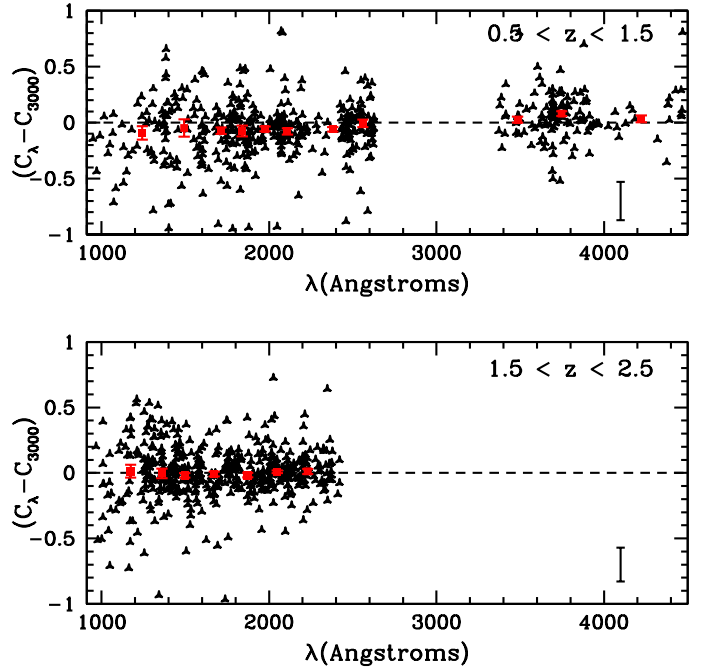}
\caption{The same as Figure~\ref{fig:Rdiff}, except we plot the
  difference in concentration. 
\label{fig:ConcDiff}}
\end{figure}

\begin{figure}[t]
\epsscale{1.3}
\plotone{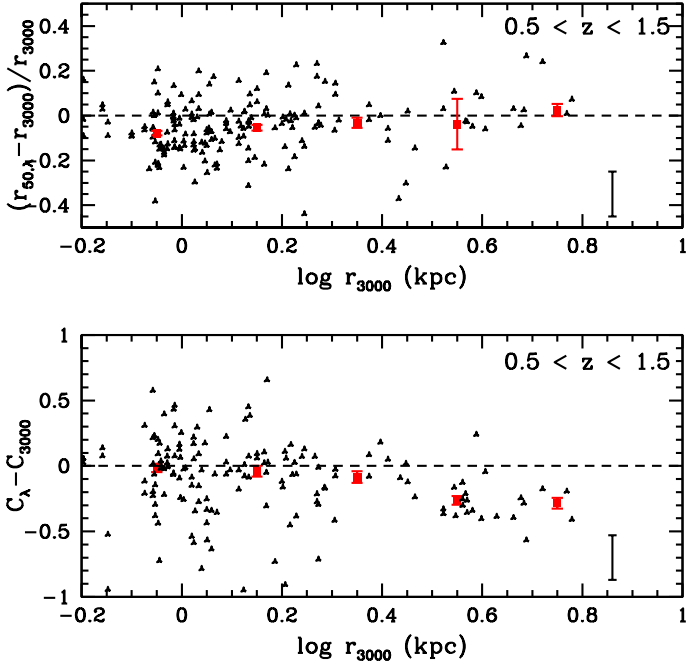}
\caption{The difference in $r_{50}$ (top panel) and
  concentration (bottom panel) as a function of $r_{3000}$
  for a sample of $0.5 < z < 1.5$ galaxies with $M_{1500}<-17$.
  Differences are taken between far-UV wavelengths,
  $\lambda_r<1800$~\AA, and the filter nearest to rest-frame
  $3000$~\AA.  Black triangles indicate individual measurements, while red
  square points indicate the median differential radius in bins of
  $\Delta {\rm log}\,r_{3000}=0.2$.  Error bars for the medians are derived from 1000
  bootstrap simulations and a typical error bar for the individual measurements is given in the lower right corner.  The fractional difference in $r_{50}$ is independent of size, while large galaxies are systematically less
  concentrated in the far-UV than the near-UV.  This is likely due to the
  presence of bulges in many of the larger objects, which are faint at
  the bluest wavelengths.
  \label{fig:Rdiff_bothr}}
\end{figure}

\begin{figure}[t]
\epsscale{1.3}
\plotone{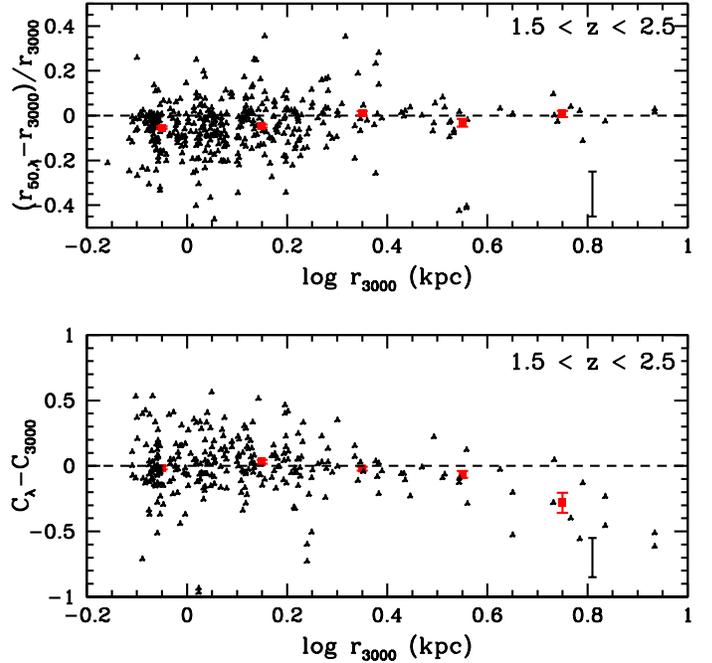}
\caption{Same as Figure~\ref{fig:Rdiff_bothr}, but for the $1.5<z<2.5$
  sample.  There is no evidence for a systematic decrease in
  concentration except for galaxies with $r_{3000}\gtrsim4$~kpc.
\label{fig:Rdiff_bothr_hz}}
\end{figure}

\section{WAVELENGTH DEPENDENCE OF $\lowercase{r_{50}}$ AND CONCENTRATION}
\label{sec:radresult}

Ultraviolet light in the rest frame of galaxies will be dominated by
recent ($\lesssim100$~Myr) star formation, but observing bluer UV
light allows us to distinguish the youngest ($\sim10$~Myr) star
formation, as well as regions least extincted by dust.  By analyzing
the wavelength dependence of simple structural diagnostics, such as
$r_{50}$ and concentration, we can characterize the spatial variation
of these changes.  In Table~\ref{tab:MainSample}, we give the
position, photometric redshift, $M_{1500}$, $r_{50}$ at $3000$~\AA\ ($r_{3000}$), and $C$ at
$3000$~\AA\ ($C_{3000}$) for each galaxy in the $0.5 < z < 1.5$
sample.

In the top panel of Figure~\ref{fig:Rdiff}, we show
how $r_{50}$ depends upon rest-frame wavelength in our sample of $0.5 < z
< 1.5$ galaxies.  For each galaxy, we plot the fractional difference
in $r_{50}$ relative to the filter nearest to rest-frame
$3000$~\AA.  This choice of wavelength ensures that the reference
measurement is taken in the observed-frame optical UDF images, in
which all of our galaxies are detected at high S/N.  The reader should
bear in mind that our observations use broadband filters, with typical
widths at $z\sim 1$ ranging from $\Delta \lambda_r \sim 200$~\AA\ at
$1000$~\AA\ to $\Delta \lambda_r \sim 1000$~\AA\ at $4000$~\AA.
Measurements near $912$~\AA, for example, will include light both
blueward and redward of the Lyman Break.

As shown in Figure~\ref{fig:Rdiff}, the median $r_{50}$
decreases at $\lambda<1800$~\AA\ for galaxies in both redshift
intervals.  When we consider only points bluer than this wavelength,
the median change in $r_{50}$ is 5.5\%$\pm 0.9$\%, while averaging over all
$\lambda<3000$~\AA\ yields a change of 3.0\%$\pm 0.5$\%.  At $z\sim2$,
the same measurements give a decrease in $r_{50}$ of 3.7\%$\pm0.6$\% and
3.2\%$\pm0.3$\%.  The concentration, shown in
Figure~\ref{fig:ConcDiff}, also decreases at $z\sim1$ for
$\lambda<1800$~\AA, with $\Delta C=-0.07\pm0.02$ at $z\sim 1$ ($\Delta
C=-0.06\pm0.01$ for $\lambda<3000$~\AA).  
This is not consistent with the change in concentration seen at
$z\sim2$ ($\Delta C=-0.01\pm0.01$ for $\lambda<1800$~\AA), suggesting
that the wavelength dependence of concentration is more substantial at
low redshift.

We also plot the fractional difference in $r_{50}$ and concentration as a
function of $r_{3000}$ (see Figure~\ref{fig:Rdiff_bothr}) for
all measurements at $\lambda_r<1800$~\AA.  In addition to ensuring
that the changes in morphology in the UVIS images are not due to the
point spread function, this also helps us isolate the galaxies that
are dominating the average trends shown in Figures~\ref{fig:Rdiff} and
\ref{fig:ConcDiff}.  While the decrease of $r_{50}$ in the FUV is
approximately constant as a function of $r_{3000}$, concentration only
decreases for galaxies with $r_{3000}\gtrsim2$~kpc.  At $z\sim2$, this
effect is still present (Figure~\ref{fig:Rdiff_bothr_hz}), but only
for larger galaxies ($r_{3000}\gtrsim4$~kpc).

To better understand the origin of the changes in concentration, we
used the \citet{VisMorph} catalog to obtain $H$-band visual
classifications for $111$ galaxies in the $0.5 < z < 1.5$ sample with
$H<24.5$, all with at least four reliable classifications.  We define
disky galaxies to be those with spheroidicity $>0.5$, where
classifiers identified the galaxy as being disk-dominated.  By
contrast, spheroidal galaxies are those that were considered to be
bulge-dominated in the rest-frame optical, or spheroidicity $<0.5$.
For spheroidal galaxies, the median change in concentration at
$\lambda<1800$~\AA\ is consistent with zero, ($C-C_{3000}=-0.02 \pm
0.03$), while disky galaxies exhibit a significant drop in
concentration over the same wavelength range, with $C-C_{3000}=-0.20
\pm 0.05$.  This result suggests that the decrease in concentration is
occurring primarily in disk-like galaxies.

\begin{figure}[t]
\epsscale{1.3}
\plotone{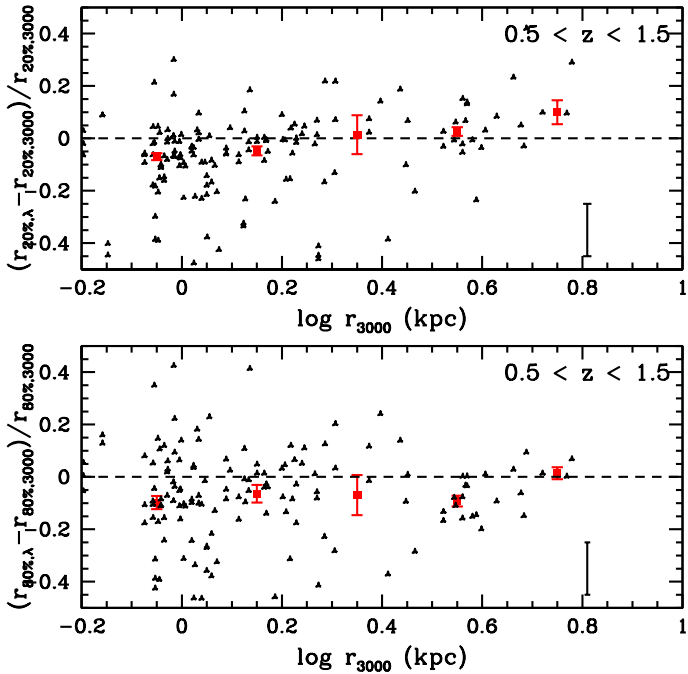}
\caption{The difference in the $20$\%\ light radius (top panel) and
  the $80$\%\ light radius (bottom panel) as a function of $r_{3000}$
  for a sample of $0.5 < z < 1.5$ galaxies with $M_{1500}<-17$.  Data
  points are generated in the same way as in
  Figure~\ref{fig:Rdiff_bothr} and a typical error bar for the individual
  measurements is given in the lower right corner.  The fractional difference in the $20$\%\ light radius rises at large $r_{3000}$, sugesting that the central profiles of large galaxies are
  flattening in the far-UV.  This is likely due to the  presence of
  bulges or proto-bulges in many of the larger objects at $z\sim1$.
  \label{fig:Rdiff_bothr20}}
\end{figure}

We can isolate the cause of this change even further by separately
analyzing the wavelength dependence of $r_{20\%}$ and $r_{80\%}$ for
the galaxies in our sample.  Note that the difference in concetration
between two wavelengths can be expressed as $\Delta C$, where
\begin{equation}
\Delta
C=C_{\lambda_1}-C_{\lambda_2}=5\mathrm {log}\left[\frac{r_{80\%,\lambda_1}}{r_{80\%,\lambda_2}}\frac{r_{20\%,\lambda_2}}{r_{20\%,\lambda_1}}\right].
\end{equation}
\label{eq:diffconc}
A decrease in concentration in the FUV can occur due to a flattening
of the inner part of the light profile, $r_{20\%,1500}>r_{20\%,3000}$,
or a truncation of the wings, $r_{80\%,1500}<r_{80\%,3000}$, or some
combination of both.  In Figure~\ref{fig:Rdiff_bothr20}, we show the
difference of these parameters in the FUV as a function of $r_{3000}$.
There is no evidence for a truncation of the outer light profile in
the FUV for large galaxies.  However, we do find evidence for an
increase in the inner radius, with $\Delta
r_{20\%}/r_{20\%,3000}=0.12\pm0.03$ for galaxies with $r_{3000}>2$~kpc
compared to galaxies with $r_{3000}<2$~kpc.  This suggests that the
trend in Figure~\ref{fig:Rdiff_bothr} is driven by a flattening of the
central portion of the light profile, likely due to a decreasing
contribution from the bulge of disk-like galaxies further in the FUV.

\section{DISCUSSION}
\label{sec:discussion}

Previous studies of star-forming galaxies at $0.5<z<1.5$ in the
CANDELS survey \citep{Wuyts12} revealed that the youngest stellar
populations at $0.5 < z < 1.5$ tended to be concentrated in clumps
near the effective radius (approximately equivalent to
$r_{50}$).  In a simple 1 Gyr constant star formation model, they found
that stars $<10$~Myr old will contribute $\sim 60$\% of the FUV light
and stars $<100$~Myr old will contribute $>90$\%.  Therefore, we
expect that young star-forming clumps, when present, will tend to set
the physical scale on which both FUV and NUV emission are observed and
$r_{50}$ should be approximately constant across this rest-frame
wavelength range.

However, far-UV observations of local Sa-Sb galaxies do reveal
differences between the $3000$~\AA\ and $1500$~\AA\ light profiles; in
particular, they find that galaxies of type later than S0 exhibit a
drop in concentration as one observes further into the FUV
\citep{TM07}.  They attribute this change primarily to the diminished
brightness of bulges at shorter wavelengths.  Although the fraction of
bulge-dominated galaxies decreases with redshift, we
still expect $\sim60$\%\ of our galaxies to be bulge-dominated at
$z\sim1$ \citep{Bruce12}.

\begin{figure}[t]
\plotone{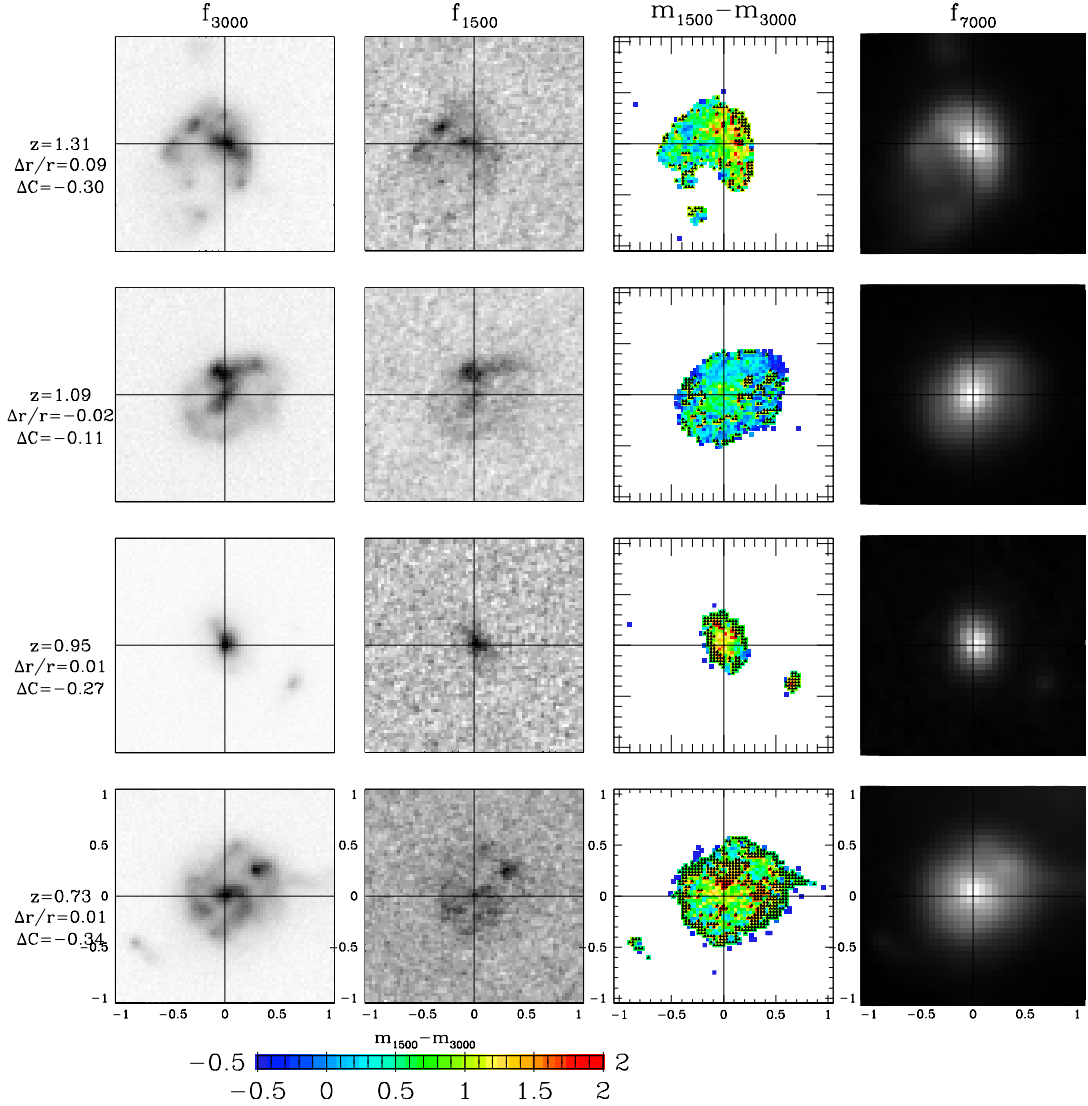}
\caption{Panels demonstrating the change in morphology of $0.5<z<1.5$
  galaxies between rest-frame $1500$~\AA\ and $3000$~\AA.  We show
  2\arcsec\ postage stamps of rest-frame $3000$~\AA\ (nearest UDF
  optical image), rest-frame $1500$~\AA\ (nearest UVIS image),
  $m_{1500}-m_{3000}$, and rest-frame $7000$~\AA\ (nearest HUDF12 image).  The greyscale images are all scaled
  relative to the minimum and maximum in each panel and the
  colorscale is given at the bottom.  When a pixel is not detected in
  the UVIS cutout, we color it with the $1-\sigma$ lower limit and
  overplot a small black triangle.  To the left of each row, we give the
  galaxy redshift, difference in $r_{50}$ between $1500$~\AA\
  and $3000$~\AA, and difference in concentration between $1500$~\AA\
  and $3000$~\AA.
  \label{fig:Color3v15}}
\end{figure}

Overall, our results are consistent with these expectations, although
we do observe a small decrease in $r_{50}$ ($\sim 5$\%) in the FUV for
samples at both $0.5<z<1.5$ and $1.5<z<2.5$.  The cause of this
decrease is not clear, but it is independent of galaxy size.  We also
observe a decrease in concentration in the FUV, consistent with
results at low redshift.  It is only marginal for the sample as a
whole ($\Delta C\simeq0.05$, Figure~\ref{fig:ConcDiff}), but the
largest galaxies ($r_{3000}>2$~kpc) exhibit a drop of $\Delta C \simeq
-0.3$, which we find to be due to a flattening of the central portion
of the light profile for $\lambda<1800$~\AA.  A few illustrative
examples are shown in Figure~\ref{fig:Color3v15}, where we plot the
pixel-by-pixel color maps of four $0.5<z<1.5$ galaxies between
rest-frame $1500$~\AA\ and $3000$~\AA.  We also show NIR cutouts from
HUDF12 for comparison.  While the majority of the UV emission is blue,
with $m_{1500}-m_{3000} \sim 0 - 1$, the region near the rest-optical
centroid tend to be redder than the rest of the galaxy.  This is
likely due to the presence of a bulge or proto-bulge near the center
of the galaxy with older stellar populations and/or more dust than the
rest of the galaxy.

To summarize, we find that a $1500$~\AA\ luminosity-limited sample of
galaxies at $0.5<z<1.5$ is both smaller ($\sim 5$\%) and less
concentrated ($\Delta C\simeq0.05$) at $1500$~\AA\ compared to
$3000$~\AA.  While the wavelength dependence of $r_{50}$ is
independent $r_{3000}$ at all redshifts studied, the decrease in
concentration is more substantial for galaxies with $r\gtrsim2$~kpc at
$z\sim1$.  At $z\sim2$, concentration is approximately constant across
the rest-UV for all but the largest galaxies ($r\gtrsim4$~kpc).  While
we have painted a broad picture of the structural properties of
star-forming galaxies in the FUV, a careful analysis of the spatial,
size, and color distribution of star-forming clumps is underway (de
Mello et al, in prep) and should provide us with a more detailed
picture of the star formation in these galaxies.

\acknowledgments

Support for program number HST-GO-12534 was provided by NASA through a grant from the Space Telescope Science Institute, which is operated by the Association of Universities for Research in Astronomy, Inc., under NASA contract NAS 5-26555.

\bibliographystyle{apj}                    

\bibliography{apj-jour,Bond_Resubmit}  

\newpage

\begin{longtable}{lcccccc}
\tablecaption{Galaxy Samples\label{tab:MainSample}}
\tablewidth{0pt}
\tablehead{
\colhead{Number}
&\colhead{$\alpha$}
&\colhead{$\delta$}
&\colhead{$z$}
&\colhead{$M_{1500}$}
&\colhead{$r_{3000}$ (")}
&\colhead{$C_{3000}$}
\\
&
&
&
&
}
$1267$ &$3$:$32$:$38.522$ &$-27$:$48$:$38.327$ &$1.341$ &$-17.05 \pm 0.12$ &$0.119 \pm 0.005$ &$2.22 \pm 0.08$  \\
$1269$ &$3$:$32$:$38.608$ &$-27$:$48$:$37.577$ &$1.425$ &$-17.90 \pm 0.09$ &$0.157 \pm 0.007$ &$2.50 \pm 0.09$  \\
$1478$ &$3$:$32$:$38.397$ &$-27$:$48$:$28.997$ &$1.489$ &$-18.20 \pm 0.08$ &$0.441 \pm 0.016$ &$2.55 \pm 0.12$  \\
$1481$ &$3$:$32$:$39.982$ &$-27$:$48$:$30.610$ &$0.959$ &$-17.10 \pm 0.12$ &$0.189 \pm 0.007$ &$2.56 \pm 0.10$  \\
$1531$ &$3$:$32$:$37.751$ &$-27$:$48$:$30.170$ &$1.496$ &$-17.20 \pm 0.12$ &$0.219 \pm 0.019$ &$3.00 \pm 0.14$  \\
$1626$ &$3$:$32$:$38.352$ &$-27$:$48$:$25.728$ &$1.458$ &$-18.30 \pm 0.08$ &$0.517 \pm 0.019$ &$2.01 \pm 0.11$  \\
$1668$ &$3$:$32$:$40.482$ &$-27$:$48$:$26.558$ &$1.167$ &$-18.47 \pm 0.07$ &$0.129 \pm 0.003$ &$2.67 \pm 0.07$  \\
$1693$ &$3$:$32$:$37.877$ &$-27$:$48$:$26.719$ &$1.120$ &$-17.01 \pm 0.13$ &$0.130 \pm 0.006$ &$2.67 \pm 0.09$  \\
$1752$ &$3$:$32$:$40.923$ &$-27$:$48$:$24.006$ &$1.470$ &$-18.81 \pm 0.05$ &$0.098 \pm 0.002$ &$3.82 \pm 0.08$  \\
$1829$ &$3$:$32$:$40.927$ &$-27$:$48$:$23.646$ &$1.298$ &$-18.36 \pm 0.07$ &$0.165 \pm 0.004$ &$2.73 \pm 0.07$  \\
$1960$ &$3$:$32$:$35.964$ &$-27$:$48$:$11.906$ &$0.605$ &$-17.07 \pm 0.12$ &$0.129 \pm 0.001$ &$3.76 \pm 0.04$  \\
$2322$ &$3$:$32$:$39.111$ &$-27$:$48$:$01.844$ &$1.270$ &$-18.27 \pm 0.07$ &$0.102 \pm 0.003$ &$2.49 \pm 0.07$  \\
$2333$ &$3$:$32$:$36.853$ &$-27$:$48$:$13.103$ &$1.310$ &$-17.32 \pm 0.11$ &$0.237 \pm 0.011$ &$2.39 \pm 0.10$  \\
$2461$ &$3$:$32$:$41.587$ &$-27$:$48$:$08.252$ &$1.123$ &$-17.18 \pm 0.12$ &$0.150 \pm 0.008$ &$3.04 \pm 0.11$  \\
$2763$ &$3$:$32$:$36.613$ &$-27$:$48$:$01.254$ &$1.492$ &$-18.64 \pm 0.06$ &$0.335 \pm 0.009$ &$3.03 \pm 0.11$  \\
$2934$ &$3$:$32$:$37.696$ &$-27$:$48$:$02.420$ &$1.143$ &$-17.37 \pm 0.11$ &$0.110 \pm 0.004$ &$2.45 \pm 0.07$  \\
$2998$ &$3$:$32$:$36.287$ &$-27$:$47$:$55.285$ &$0.713$ &$-17.73 \pm 0.10$ &$0.361 \pm 0.005$ &$2.33 \pm 0.07$  \\
$3031$ &$3$:$32$:$36.387$ &$-27$:$47$:$58.585$ &$1.439$ &$-18.97 \pm 0.05$ &$0.474 \pm 0.015$ &$2.55 \pm 0.11$  \\
$3123$ &$3$:$32$:$42.948$ &$-27$:$47$:$55.134$ &$1.065$ &$-18.08 \pm 0.08$ &$0.185 \pm 0.004$ &$2.95 \pm 0.08$  \\
$3180$ &$3$:$32$:$37.879$ &$-27$:$47$:$51.079$ &$0.768$ &$-18.91 \pm 0.05$ &$0.336 \pm 0.006$ &$2.56 \pm 0.10$  \\
$3243$ &$3$:$32$:$38.648$ &$-27$:$47$:$56.206$ &$1.355$ &$-17.59 \pm 0.11$ &$0.367 \pm 0.016$ &$2.48 \pm 0.12$  \\
$3257$ &$3$:$32$:$38.955$ &$-27$:$47$:$55.095$ &$1.379$ &$-17.99 \pm 0.09$ &$0.501 \pm 0.026$ &$2.16 \pm 0.12$  \\
$3270$ &$3$:$32$:$38.670$ &$-27$:$47$:$55.696$ &$1.229$ &$-17.86 \pm 0.09$ &$0.255 \pm 0.011$ &$3.07 \pm 0.12$  \\
$3349$ &$3$:$32$:$41.675$ &$-27$:$47$:$50.462$ &$0.668$ &$-17.29 \pm 0.12$ &$0.500 \pm 0.008$ &$2.63 \pm 0.09$  \\
$3372$ &$3$:$32$:$42.247$ &$-27$:$47$:$46.139$ &$0.794$ &$-18.69 \pm 0.07$ &$0.734 \pm 0.004$ &$2.68 \pm 0.08$  \\
$3373$ &$3$:$32$:$40.049$ &$-27$:$47$:$51.790$ &$0.995$ &$-18.60 \pm 0.07$ &$0.449 \pm 0.007$ &$2.02 \pm 0.08$  \\
$3613$ &$3$:$32$:$37.632$ &$-27$:$47$:$44.300$ &$1.097$ &$-19.62 \pm 0.02$ &$0.410 \pm 0.004$ &$2.76 \pm 0.07$  \\
$3655$ &$3$:$32$:$37.567$ &$-27$:$47$:$50.181$ &$1.327$ &$-17.03 \pm 0.12$ &$0.110 \pm 0.005$ &$2.45 \pm 0.08$  \\
$3677$ &$3$:$32$:$37.309$ &$-27$:$47$:$29.362$ &$0.669$ &$-18.35 \pm 0.07$ &$0.270 \pm 0.001$ &$3.37 \pm 0.06$  \\
$3752$ &$3$:$32$:$43.246$ &$-27$:$47$:$44.003$ &$1.440$ &$-18.28 \pm 0.07$ &$0.154 \pm 0.008$ &$3.04 \pm 0.10$  \\
$3799$ &$3$:$32$:$38.353$ &$-27$:$47$:$44.418$ &$1.442$ &$-18.31 \pm 0.08$ &$0.710 \pm 0.031$ &$2.14 \pm 0.13$  \\
$3977$ &$3$:$32$:$37.397$ &$-27$:$47$:$41.601$ &$1.095$ &$-19.26 \pm 0.04$ &$0.555 \pm 0.007$ &$2.65 \pm 0.08$  \\
$4052$ &$3$:$32$:$40.632$ &$-27$:$47$:$39.997$ &$1.040$ &$-17.50 \pm 0.11$ &$0.240 \pm 0.008$ &$2.94 \pm 0.11$  \\
$4094$ &$3$:$32$:$37.594$ &$-27$:$47$:$39.531$ &$0.663$ &$-17.96 \pm 0.09$ &$0.136 \pm 0.005$ &$3.98 \pm 0.12$  \\
$4142$ &$3$:$32$:$44.197$ &$-27$:$47$:$33.527$ &$0.737$ &$-19.28 \pm 0.03$ &$0.240 \pm 0.001$ &$2.79 \pm 0.02$  \\
$4253$ &$3$:$32$:$39.888$ &$-27$:$47$:$38.261$ &$1.049$ &$-18.92 \pm 0.05$ &$0.459 \pm 0.010$ &$2.39 \pm 0.10$  \\
$4332$ &$3$:$32$:$33.456$ &$-27$:$47$:$39.512$ &$1.403$ &$-17.92 \pm 0.09$ &$0.131 \pm 0.004$ &$2.57 \pm 0.07$  \\
$4396$ &$3$:$32$:$35.796$ &$-27$:$47$:$34.736$ &$1.223$ &$-17.97 \pm 0.09$ &$0.145 \pm 0.005$ &$3.19 \pm 0.10$  \\
$4438$ &$3$:$32$:$33.031$ &$-27$:$47$:$30.633$ &$0.977$ &$-18.82 \pm 0.06$ &$0.700 \pm 0.007$ &$2.31 \pm 0.09$  \\
$4458$ &$3$:$32$:$36.054$ &$-27$:$47$:$37.796$ &$1.340$ &$-21.47 \pm 0.06$ &$0.132 \pm 0.017$ &$1.13 \pm 0.13$  \\
$4481$ &$3$:$32$:$39.264$ &$-27$:$47$:$36.704$ &$1.367$ &$-18.87 \pm 0.05$ &$0.187 \pm 0.004$ &$2.87 \pm 0.07$  \\
$4491$ &$3$:$32$:$40.216$ &$-27$:$47$:$32.979$ &$1.095$ &$-19.53 \pm 0.03$ &$0.578 \pm 0.006$ &$2.69 \pm 0.08$  \\
$4587$ &$3$:$32$:$40.673$ &$-27$:$47$:$30.997$ &$0.667$ &$-17.02 \pm 0.13$ &$0.266 \pm 0.001$ &$2.75 \pm 0.04$  \\
$4591$ &$3$:$32$:$41.120$ &$-27$:$47$:$34.595$ &$1.009$ &$-18.13 \pm 0.08$ &$0.410 \pm 0.009$ &$2.87 \pm 0.10$  \\
$4616$ &$3$:$32$:$42.737$ &$-27$:$47$:$33.986$ &$1.427$ &$-17.63 \pm 0.11$ &$0.435 \pm 0.022$ &$2.46 \pm 0.13$  \\
$4662$ &$3$:$32$:$39.490$ &$-27$:$47$:$34.663$ &$1.098$ &$-17.69 \pm 0.10$ &$0.221 \pm 0.009$ &$3.10 \pm 0.11$  \\
$4767$ &$3$:$32$:$40.607$ &$-27$:$47$:$30.247$ &$0.669$ &$-17.01 \pm 0.15$ &$1.384 \pm 0.010$ &$1.15 \pm 0.09$  \\
$4816$ &$3$:$32$:$44.165$ &$-27$:$47$:$29.447$ &$1.220$ &$-18.37 \pm 0.07$ &$0.198 \pm 0.004$ &$3.29 \pm 0.09$  \\
$4835$ &$3$:$32$:$34.867$ &$-27$:$47$:$30.689$ &$1.317$ &$-18.02 \pm 0.08$ &$0.188 \pm 0.006$ &$2.97 \pm 0.10$  \\
$4849$ &$3$:$32$:$39.321$ &$-27$:$47$:$32.174$ &$1.379$ &$-17.42 \pm 0.11$ &$0.165 \pm 0.009$ &$2.50 \pm 0.10$  \\
$4976$ &$3$:$32$:$34.693$ &$-27$:$47$:$28.019$ &$1.466$ &$-18.93 \pm 0.05$ &$0.186 \pm 0.004$ &$3.12 \pm 0.10$  \\
$4981$ &$3$:$32$:$34.659$ &$-27$:$47$:$28.019$ &$1.438$ &$-19.06 \pm 0.04$ &$0.157 \pm 0.003$ &$2.91 \pm 0.08$  \\
$4999$ &$3$:$32$:$36.352$ &$-27$:$47$:$27.985$ &$1.378$ &$-18.82 \pm 0.05$ &$0.215 \pm 0.008$ &$3.48 \pm 0.11$  \\
$5115$ &$3$:$32$:$32.651$ &$-27$:$47$:$27.333$ &$1.355$ &$-17.33 \pm 0.11$ &$0.136 \pm 0.011$ &$3.61 \pm 0.13$  \\
$5187$ &$3$:$32$:$44.355$ &$-27$:$47$:$23.776$ &$0.953$ &$-17.66 \pm 0.10$ &$0.163 \pm 0.002$ &$2.77 \pm 0.06$  \\
$5190$ &$3$:$32$:$34.808$ &$-27$:$47$:$21.839$ &$1.316$ &$-19.59 \pm 0.03$ &$0.443 \pm 0.005$ &$2.49 \pm 0.08$  \\
$5216$ &$3$:$32$:$34.261$ &$-27$:$47$:$24.090$ &$1.098$ &$-18.16 \pm 0.09$ &$0.570 \pm 0.018$ &$1.73 \pm 0.11$  \\
$5268$ &$3$:$32$:$40.324$ &$-27$:$47$:$22.809$ &$0.619$ &$-17.85 \pm 0.10$ &$0.336 \pm 0.003$ &$2.26 \pm 0.06$  \\
$5388$ &$3$:$32$:$41.857$ &$-27$:$47$:$21.901$ &$1.325$ &$-18.32 \pm 0.08$ &$0.368 \pm 0.010$ &$2.76 \pm 0.10$  \\
$5417$ &$3$:$32$:$39.881$ &$-27$:$47$:$15.011$ &$1.095$ &$-20.17 \pm 0.01$ &$0.935 \pm 0.005$ &$2.02 \pm 0.08$  \\
$5497$ &$3$:$32$:$37.763$ &$-27$:$47$:$21.200$ &$1.092$ &$-17.28 \pm 0.12$ &$0.284 \pm 0.011$ &$2.31 \pm 0.10$  \\
$5569$ &$3$:$32$:$35.328$ &$-27$:$47$:$18.478$ &$0.674$ &$-17.47 \pm 0.11$ &$0.273 \pm 0.003$ &$2.76 \pm 0.06$  \\
$5658$ &$3$:$32$:$41.766$ &$-27$:$47$:$16.831$ &$1.096$ &$-18.74 \pm 0.06$ &$0.272 \pm 0.007$ &$3.78 \pm 0.11$  \\
$5661$ &$3$:$32$:$31.776$ &$-27$:$47$:$20.194$ &$1.412$ &$-17.55 \pm 0.10$ &$0.178 \pm 0.009$ &$2.37 \pm 0.10$  \\
$5694$ &$3$:$32$:$43.475$ &$-27$:$47$:$12.921$ &$1.095$ &$-19.01 \pm 0.05$ &$0.630 \pm 0.012$ &$2.55 \pm 0.11$  \\
$5709$ &$3$:$32$:$35.667$ &$-27$:$47$:$19.137$ &$1.221$ &$-17.28 \pm 0.11$ &$0.104 \pm 0.005$ &$2.49 \pm 0.08$  \\
$5753$ &$3$:$32$:$32.328$ &$-27$:$47$:$18.364$ &$1.027$ &$-17.17 \pm 0.12$ &$0.194 \pm 0.010$ &$3.06 \pm 0.11$  \\
$5896$ &$3$:$32$:$36.958$ &$-27$:$47$:$15.773$ &$1.123$ &$-17.59 \pm 0.10$ &$0.141 \pm 0.004$ &$3.08 \pm 0.09$  \\
$5898$ &$3$:$32$:$31.397$ &$-27$:$47$:$13.085$ &$1.479$ &$-18.26 \pm 0.08$ &$0.351 \pm 0.014$ &$3.06 \pm 0.12$  \\
$5922$ &$3$:$32$:$31.568$ &$-27$:$47$:$11.164$ &$1.008$ &$-18.10 \pm 0.09$ &$0.522 \pm 0.017$ &$2.60 \pm 0.12$  \\
$5959$ &$3$:$32$:$39.381$ &$-27$:$47$:$14.263$ &$1.109$ &$-18.43 \pm 0.07$ &$0.122 \pm 0.002$ &$2.42 \pm 0.04$  \\
$5975$ &$3$:$32$:$41.692$ &$-27$:$47$:$13.562$ &$1.450$ &$-18.58 \pm 0.06$ &$0.204 \pm 0.006$ &$2.71 \pm 0.09$  \\
$5989$ &$3$:$32$:$38.633$ &$-27$:$47$:$11.356$ &$1.134$ &$-18.59 \pm 0.07$ &$0.395 \pm 0.007$ &$2.44 \pm 0.08$  \\
$5995$ &$3$:$32$:$42.662$ &$-27$:$47$:$13.106$ &$0.968$ &$-17.99 \pm 0.09$ &$0.217 \pm 0.005$ &$2.71 \pm 0.09$  \\
$6022$ &$3$:$32$:$35.310$ &$-27$:$47$:$13.588$ &$1.392$ &$-18.30 \pm 0.07$ &$0.116 \pm 0.002$ &$2.73 \pm 0.06$  \\
$6027$ &$3$:$32$:$39.637$ &$-27$:$47$:$09.132$ &$1.317$ &$-17.07 \pm 0.12$ &$0.129 \pm 0.004$ &$3.93 \pm 0.12$  \\
$6079$ &$3$:$32$:$33.458$ &$-27$:$47$:$12.392$ &$1.298$ &$-18.02 \pm 0.09$ &$0.226 \pm 0.006$ &$2.84 \pm 0.09$  \\
$6107$ &$3$:$32$:$39.838$ &$-27$:$47$:$11.861$ &$1.081$ &$-17.49 \pm 0.11$ &$0.271 \pm 0.012$ &$2.36 \pm 0.10$  \\
$6118$ &$3$:$32$:$43.498$ &$-27$:$47$:$11.931$ &$1.090$ &$-17.86 \pm 0.09$ &$0.187 \pm 0.007$ &$2.28 \pm 0.13$  \\
$6153$ &$3$:$32$:$34.941$ &$-27$:$47$:$10.169$ &$1.380$ &$-17.18 \pm 0.12$ &$0.126 \pm 0.009$ &$2.41 \pm 0.12$  \\
$6187$ &$3$:$32$:$37.480$ &$-27$:$47$:$10.491$ &$0.980$ &$-17.16 \pm 0.12$ &$0.230 \pm 0.011$ &$2.54 \pm 0.10$  \\
$6196$ &$3$:$32$:$38.513$ &$-27$:$47$:$10.667$ &$0.990$ &$-17.73 \pm 0.10$ &$0.131 \pm 0.003$ &$2.56 \pm 0.06$  \\
$6235$ &$3$:$32$:$34.907$ &$-27$:$47$:$10.439$ &$1.380$ &$-17.63 \pm 0.10$ &$0.150 \pm 0.011$ &$2.68 \pm 0.13$  \\
$6583$ &$3$:$32$:$36.447$ &$-27$:$47$:$05.094$ &$1.224$ &$-17.33 \pm 0.11$ &$0.126 \pm 0.006$ &$2.46 \pm 0.08$  \\
$6627$ &$3$:$32$:$35.384$ &$-27$:$47$:$04.258$ &$0.989$ &$-17.46 \pm 0.11$ &$0.122 \pm 0.004$ &$2.51 \pm 0.07$  \\
$6645$ &$3$:$32$:$33.176$ &$-27$:$47$:$03.842$ &$1.326$ &$-18.18 \pm 0.08$ &$0.498 \pm 0.017$ &$2.21 \pm 0.11$  \\
$6673$ &$3$:$32$:$34.767$ &$-27$:$47$:$04.379$ &$1.272$ &$-17.08 \pm 0.12$ &$0.116 \pm 0.006$ &$2.39 \pm 0.08$  \\
$6710$ &$3$:$32$:$46.235$ &$-27$:$47$:$01.682$ &$0.980$ &$-17.64 \pm 0.10$ &$0.244 \pm 0.009$ &$3.57 \pm 0.12$  \\
$6754$ &$3$:$32$:$31.292$ &$-27$:$47$:$02.375$ &$1.093$ &$-17.82 \pm 0.09$ &$0.148 \pm 0.004$ &$2.92 \pm 0.09$  \\
$6782$ &$3$:$32$:$32.506$ &$-27$:$47$:$01.983$ &$1.446$ &$-17.81 \pm 0.09$ &$0.156 \pm 0.006$ &$2.38 \pm 0.09$  \\
$6785$ &$3$:$32$:$45.966$ &$-27$:$46$:$57.754$ &$1.331$ &$-19.24 \pm 0.05$ &$0.972 \pm 0.013$ &$2.38 \pm 0.10$  \\
$6821$ &$3$:$32$:$42.774$ &$-27$:$46$:$59.066$ &$1.119$ &$-19.30 \pm 0.04$ &$0.445 \pm 0.004$ &$2.12 \pm 0.07$  \\
$6846$ &$3$:$32$:$44.298$ &$-27$:$47$:$00.016$ &$1.143$ &$-17.78 \pm 0.09$ &$0.108 \pm 0.003$ &$2.86 \pm 0.07$  \\
$6853$ &$3$:$32$:$36.435$ &$-27$:$46$:$58.344$ &$0.852$ &$-18.51 \pm 0.06$ &$0.150 \pm 0.002$ &$2.87 \pm 0.05$  \\
$6862$ &$3$:$32$:$41.422$ &$-27$:$46$:$51.453$ &$0.620$ &$-17.49 \pm 0.11$ &$0.584 \pm 0.004$ &$2.56 \pm 0.08$  \\
$6886$ &$3$:$32$:$43.497$ &$-27$:$46$:$59.091$ &$1.400$ &$-18.23 \pm 0.08$ &$0.185 \pm 0.006$ &$3.20 \pm 0.10$  \\
$6894$ &$3$:$32$:$31.132$ &$-27$:$47$:$00.305$ &$1.160$ &$-17.00 \pm 0.13$ &$0.122 \pm 0.005$ &$2.52 \pm 0.08$  \\
$6922$ &$3$:$32$:$36.178$ &$-27$:$46$:$57.265$ &$1.315$ &$-17.72 \pm 0.11$ &$0.808 \pm 0.040$ &$2.86 \pm 0.14$  \\
$6933$ &$3$:$32$:$33.429$ &$-27$:$46$:$50.462$ &$0.735$ &$-18.21 \pm 0.08$ &$0.547 \pm 0.006$ &$2.41 \pm 0.09$  \\
$6939$ &$3$:$32$:$46.068$ &$-27$:$46$:$58.713$ &$1.050$ &$-17.39 \pm 0.11$ &$0.345 \pm 0.011$ &$2.03 \pm 0.10$  \\
$6949$ &$3$:$32$:$44.239$ &$-27$:$47$:$00.136$ &$1.379$ &$-17.25 \pm 0.11$ &$0.114 \pm 0.007$ &$2.53 \pm 0.09$  \\
$6953$ &$3$:$32$:$36.668$ &$-27$:$46$:$57.684$ &$0.765$ &$-17.48 \pm 0.11$ &$0.183 \pm 0.004$ &$2.51 \pm 0.07$  \\
$6957$ &$3$:$32$:$35.063$ &$-27$:$46$:$58.738$ &$1.441$ &$-18.44 \pm 0.07$ &$0.217 \pm 0.005$ &$2.36 \pm 0.08$  \\
$6974$ &$3$:$32$:$37.954$ &$-27$:$46$:$51.919$ &$0.620$ &$-19.10 \pm 0.05$ &$0.839 \pm 0.004$ &$2.11 \pm 0.07$  \\
$7036$ &$3$:$32$:$45.681$ &$-27$:$46$:$55.206$ &$1.317$ &$-19.30 \pm 0.03$ &$0.155 \pm 0.003$ &$3.30 \pm 0.09$  \\
$7067$ &$3$:$32$:$32.443$ &$-27$:$46$:$56.403$ &$1.454$ &$-18.33 \pm 0.07$ &$0.191 \pm 0.005$ &$2.87 \pm 0.09$  \\
$7071$ &$3$:$32$:$36.442$ &$-27$:$46$:$55.134$ &$0.900$ &$-17.90 \pm 0.09$ &$0.386 \pm 0.006$ &$2.74 \pm 0.10$  \\
$7081$ &$3$:$32$:$38.958$ &$-27$:$46$:$56.325$ &$1.436$ &$-18.54 \pm 0.07$ &$0.674 \pm 0.013$ &$1.61 \pm 0.10$  \\
$7112$ &$3$:$32$:$39.815$ &$-27$:$46$:$53.531$ &$1.110$ &$-18.57 \pm 0.07$ &$0.409 \pm 0.005$ &$1.64 \pm 0.07$  \\
$7131$ &$3$:$32$:$32.459$ &$-27$:$46$:$54.033$ &$1.451$ &$-18.22 \pm 0.08$ &$0.124 \pm 0.002$ &$2.80 \pm 0.06$  \\
$7136$ &$3$:$32$:$38.990$ &$-27$:$46$:$56.715$ &$1.429$ &$-17.65 \pm 0.10$ &$0.108 \pm 0.005$ &$2.33 \pm 0.12$  \\
$7193$ &$3$:$32$:$33.429$ &$-27$:$46$:$55.382$ &$1.456$ &$-17.32 \pm 0.11$ &$0.190 \pm 0.015$ &$2.52 \pm 0.12$  \\
$7269$ &$3$:$32$:$41.890$ &$-27$:$46$:$51.271$ &$0.734$ &$-17.58 \pm 0.11$ &$0.464 \pm 0.009$ &$3.29 \pm 0.11$  \\
$7290$ &$3$:$32$:$36.928$ &$-27$:$46$:$53.993$ &$1.229$ &$-17.17 \pm 0.12$ &$0.143 \pm 0.008$ &$2.46 \pm 0.09$  \\
$7370$ &$3$:$32$:$32.712$ &$-27$:$46$:$51.543$ &$1.451$ &$-18.09 \pm 0.08$ &$0.121 \pm 0.003$ &$2.65 \pm 0.07$  \\
$7394$ &$3$:$32$:$38.990$ &$-27$:$46$:$51.045$ &$1.372$ &$-17.75 \pm 0.09$ &$0.104 \pm 0.003$ &$2.40 \pm 0.07$  \\
$7398$ &$3$:$32$:$42.319$ &$-27$:$46$:$51.088$ &$0.631$ &$-17.15 \pm 0.12$ &$0.107 \pm 0.002$ &$2.59 \pm 0.05$  \\
$7452$ &$3$:$32$:$44.189$ &$-27$:$46$:$46.967$ &$0.670$ &$-17.02 \pm 0.13$ &$0.270 \pm 0.006$ &$2.83 \pm 0.10$  \\
$7556$ &$3$:$32$:$40.781$ &$-27$:$46$:$15.757$ &$0.622$ &$-19.72 \pm 0.04$ &$1.288 \pm 0.007$ &$2.50 \pm 0.09$  \\
$7559$ &$3$:$32$:$38.100$ &$-27$:$46$:$13.849$ &$0.998$ &$-19.07 \pm 0.05$ &$0.498 \pm 0.004$ &$2.12 \pm 0.07$  \\
$7647$ &$3$:$32$:$41.425$ &$-27$:$46$:$15.093$ &$1.010$ &$-17.17 \pm 0.12$ &$0.334 \pm 0.016$ &$3.18 \pm 0.13$  \\
$7664$ &$3$:$32$:$36.954$ &$-27$:$46$:$15.563$ &$0.865$ &$-17.76 \pm 0.10$ &$0.402 \pm 0.009$ &$3.02 \pm 0.11$  \\
$7678$ &$3$:$32$:$38.986$ &$-27$:$46$:$15.225$ &$1.049$ &$-17.84 \pm 0.09$ &$0.251 \pm 0.006$ &$2.90 \pm 0.10$  \\
$7705$ &$3$:$32$:$37.563$ &$-27$:$46$:$46.761$ &$1.337$ &$-18.28 \pm 0.08$ &$0.264 \pm 0.012$ &$2.80 \pm 0.12$  \\
$7725$ &$3$:$32$:$35.078$ &$-27$:$46$:$15.658$ &$1.316$ &$-19.40 \pm 0.04$ &$0.626 \pm 0.009$ &$2.17 \pm 0.09$  \\
$7756$ &$3$:$32$:$33.926$ &$-27$:$46$:$16.861$ &$1.492$ &$-17.79 \pm 0.10$ &$0.372 \pm 0.016$ &$2.72 \pm 0.12$  \\
$7786$ &$3$:$32$:$37.065$ &$-27$:$46$:$17.122$ &$1.274$ &$-18.44 \pm 0.07$ &$0.157 \pm 0.002$ &$3.46 \pm 0.07$  \\
$7889$ &$3$:$32$:$44.274$ &$-27$:$46$:$20.116$ &$0.877$ &$-17.23 \pm 0.12$ &$0.224 \pm 0.008$ &$3.19 \pm 0.11$  \\
$7959$ &$3$:$32$:$34.758$ &$-27$:$46$:$42.929$ &$1.484$ &$-17.37 \pm 0.11$ &$0.132 \pm 0.006$ &$2.67 \pm 0.09$  \\
$7974$ &$3$:$32$:$37.728$ &$-27$:$46$:$42.620$ &$1.307$ &$-19.10 \pm 0.04$ &$0.209 \pm 0.003$ &$2.76 \pm 0.07$  \\
$7995$ &$3$:$32$:$42.253$ &$-27$:$46$:$25.289$ &$1.288$ &$-19.82 \pm 0.02$ &$0.731 \pm 0.008$ &$1.71 \pm 0.09$  \\
$8051$ &$3$:$32$:$37.425$ &$-27$:$46$:$22.491$ &$1.106$ &$-17.85 \pm 0.09$ &$0.209 \pm 0.006$ &$2.89 \pm 0.10$  \\
$8125$ &$3$:$32$:$41.514$ &$-27$:$46$:$40.353$ &$1.102$ &$-19.32 \pm 0.04$ &$0.451 \pm 0.006$ &$2.78 \pm 0.09$  \\
$8242$ &$3$:$32$:$34.954$ &$-27$:$46$:$24.329$ &$1.100$ &$-17.33 \pm 0.11$ &$0.122 \pm 0.004$ &$2.34 \pm 0.07$  \\
$8255$ &$3$:$32$:$34.627$ &$-27$:$46$:$37.799$ &$1.101$ &$-17.35 \pm 0.11$ &$0.135 \pm 0.004$ &$2.92 \pm 0.09$  \\
$8257$ &$3$:$32$:$38.596$ &$-27$:$46$:$31.277$ &$0.618$ &$-18.29 \pm 0.09$ &$0.913 \pm 0.005$ &$2.44 \pm 0.08$  \\
$8270$ &$3$:$32$:$43.971$ &$-27$:$46$:$32.448$ &$0.980$ &$-18.73 \pm 0.06$ &$0.409 \pm 0.006$ &$2.51 \pm 0.08$  \\
$8275$ &$3$:$32$:$36.451$ &$-27$:$46$:$28.164$ &$0.772$ &$-18.76 \pm 0.05$ &$0.142 \pm 0.002$ &$2.51 \pm 0.09$  \\
$8314$ &$3$:$32$:$36.661$ &$-27$:$46$:$31.014$ &$0.999$ &$-17.81 \pm 0.09$ &$0.254 \pm 0.005$ &$3.19 \pm 0.09$  \\
$8316$ &$3$:$32$:$38.300$ &$-27$:$46$:$28.458$ &$1.122$ &$-18.65 \pm 0.06$ &$0.186 \pm 0.010$ &$3.27 \pm 0.12$  \\
$8372$ &$3$:$32$:$35.619$ &$-27$:$46$:$32.847$ &$0.586$ &$-17.78 \pm 0.10$ &$0.466 \pm 0.004$ &$2.93 \pm 0.08$  \\
$8374$ &$3$:$32$:$35.391$ &$-27$:$46$:$30.507$ &$1.069$ &$-17.06 \pm 0.12$ &$0.154 \pm 0.008$ &$2.52 \pm 0.09$  \\
$8392$ &$3$:$32$:$34.186$ &$-27$:$46$:$34.620$ &$1.376$ &$-17.36 \pm 0.11$ &$0.174 \pm 0.007$ &$2.53 \pm 0.09$  \\
$8461$ &$3$:$32$:$44.618$ &$-27$:$46$:$32.174$ &$1.426$ &$-19.46 \pm 0.04$ &$0.673 \pm 0.009$ &$1.83 \pm 0.09$  \\
$8501$ &$3$:$32$:$44.747$ &$-27$:$46$:$37.303$ &$1.377$ &$-17.01 \pm 0.13$ &$0.159 \pm 0.010$ &$2.43 \pm 0.10$  \\
$8551$ &$3$:$32$:$36.403$ &$-27$:$46$:$31.375$ &$1.018$ &$-19.27 \pm 0.04$ &$0.339 \pm 0.005$ &$2.85 \pm 0.09$  \\
$8585$ &$3$:$32$:$35.485$ &$-27$:$46$:$27.297$ &$1.099$ &$-19.64 \pm 0.02$ &$0.230 \pm 0.002$ &$3.39 \pm 0.09$  \\
$8597$ &$3$:$32$:$41.860$ &$-27$:$46$:$34.471$ &$1.485$ &$-17.08 \pm 0.12$ &$0.180 \pm 0.014$ &$2.31 \pm 0.12$  \\
$8653$ &$3$:$32$:$39.220$ &$-27$:$46$:$36.104$ &$1.319$ &$-18.40 \pm 0.07$ &$0.134 \pm 0.003$ &$2.58 \pm 0.07$  \\
$8680$ &$3$:$32$:$35.465$ &$-27$:$46$:$36.987$ &$1.086$ &$-18.40 \pm 0.07$ &$0.162 \pm 0.002$ &$2.59 \pm 0.06$  \\
$8693$ &$3$:$32$:$44.358$ &$-27$:$46$:$38.866$ &$0.524$ &$-17.23 \pm 0.12$ &$0.570 \pm 0.008$ &$3.99 \pm 0.12$  \\
$8744$ &$3$:$32$:$35.192$ &$-27$:$46$:$38.968$ &$1.080$ &$-17.44 \pm 0.11$ &$0.137 \pm 0.004$ &$2.34 \pm 0.07$  \\
$8749$ &$3$:$32$:$34.857$ &$-27$:$46$:$40.469$ &$1.099$ &$-17.75 \pm 0.10$ &$0.403 \pm 0.005$ &$2.04 \pm 0.08$  \\
$8765$ &$3$:$32$:$36.562$ &$-27$:$46$:$40.614$ &$1.414$ &$-18.89 \pm 0.05$ &$0.296 \pm 0.006$ &$2.37 \pm 0.08$  \\
$8776$ &$3$:$32$:$36.589$ &$-27$:$46$:$39.714$ &$1.425$ &$-18.08 \pm 0.08$ &$0.110 \pm 0.004$ &$2.59 \pm 0.08$  \\
$8801$ &$3$:$32$:$41.510$ &$-27$:$46$:$42.183$ &$1.308$ &$-18.00 \pm 0.09$ &$0.518 \pm 0.021$ &$2.54 \pm 0.12$  \\
$8810$ &$3$:$32$:$37.266$ &$-27$:$46$:$10.342$ &$0.736$ &$-18.75 \pm 0.06$ &$0.501 \pm 0.003$ &$2.09 \pm 0.06$  \\
$8872$ &$3$:$32$:$43.038$ &$-27$:$46$:$43.974$ &$1.343$ &$-17.19 \pm 0.12$ &$0.156 \pm 0.016$ &$2.96 \pm 0.13$  \\
$8930$ &$3$:$32$:$37.500$ &$-27$:$46$:$45.351$ &$1.270$ &$-17.67 \pm 0.10$ &$0.167 \pm 0.013$ &$2.82 \pm 0.12$  \\
$8941$ &$3$:$32$:$37.644$ &$-27$:$46$:$41.810$ &$1.317$ &$-17.14 \pm 0.12$ &$0.121 \pm 0.006$ &$2.43 \pm 0.09$  \\
$9018$ &$3$:$32$:$35.296$ &$-27$:$46$:$42.328$ &$1.098$ &$-18.96 \pm 0.05$ &$0.495 \pm 0.007$ &$2.18 \pm 0.09$  \\
$9090$ &$3$:$32$:$39.350$ &$-27$:$46$:$09.673$ &$1.378$ &$-17.85 \pm 0.09$ &$0.188 \pm 0.013$ &$1.82 \pm 0.14$  \\
$9125$ &$3$:$32$:$39.915$ &$-27$:$46$:$06.911$ &$1.294$ &$-19.82 \pm 0.02$ &$0.707 \pm 0.009$ &$2.06 \pm 0.09$  \\
$9183$ &$3$:$32$:$38.446$ &$-27$:$46$:$09.527$ &$1.064$ &$-18.65 \pm 0.06$ &$0.448 \pm 0.009$ &$2.33 \pm 0.10$  \\
$9244$ &$3$:$32$:$38.760$ &$-27$:$46$:$03.496$ &$0.690$ &$-17.85 \pm 0.09$ &$0.149 \pm 0.002$ &$2.72 \pm 0.05$  \\
$9253$ &$3$:$32$:$42.813$ &$-27$:$46$:$05.635$ &$0.676$ &$-19.62 \pm 0.03$ &$0.874 \pm 0.003$ &$2.65 \pm 0.06$  \\
$9264$ &$3$:$32$:$37.192$ &$-27$:$46$:$08.062$ &$1.096$ &$-17.82 \pm 0.09$ &$0.191 \pm 0.002$ &$3.53 \pm 0.08$  \\
$9273$ &$3$:$32$:$40.162$ &$-27$:$46$:$05.680$ &$1.068$ &$-17.79 \pm 0.10$ &$0.403 \pm 0.021$ &$2.07 \pm 0.12$  \\
$9332$ &$3$:$32$:$43.120$ &$-27$:$46$:$07.553$ &$1.486$ &$-18.99 \pm 0.05$ &$0.492 \pm 0.010$ &$2.93 \pm 0.10$  \\
$9341$ &$3$:$32$:$38.365$ &$-27$:$46$:$00.618$ &$1.046$ &$-18.40 \pm 0.07$ &$0.395 \pm 0.008$ &$2.35 \pm 0.10$  \\
$9348$ &$3$:$32$:$36.871$ &$-27$:$46$:$04.013$ &$0.640$ &$-17.27 \pm 0.11$ &$0.140 \pm 0.004$ &$3.63 \pm 0.11$  \\
$9402$ &$3$:$32$:$35.056$ &$-27$:$45$:$59.698$ &$0.954$ &$-18.69 \pm 0.06$ &$0.341 \pm 0.007$ &$2.92 \pm 0.10$  \\
$9437$ &$3$:$32$:$35.806$ &$-27$:$45$:$49.046$ &$0.738$ &$-18.61 \pm 0.07$ &$0.759 \pm 0.007$ &$2.48 \pm 0.08$  \\
$9444$ &$3$:$32$:$37.298$ &$-27$:$45$:$57.892$ &$1.096$ &$-17.10 \pm 0.13$ &$0.376 \pm 0.008$ &$2.67 \pm 0.10$  \\
$9455$ &$3$:$32$:$41.334$ &$-27$:$45$:$54.454$ &$0.581$ &$-18.69 \pm 0.06$ &$0.519 \pm 0.005$ &$2.90 \pm 0.09$  \\
$9474$ &$3$:$32$:$39.467$ &$-27$:$45$:$57.073$ &$1.065$ &$-17.93 \pm 0.09$ &$0.445 \pm 0.010$ &$1.75 \pm 0.10$  \\
$9532$ &$3$:$32$:$36.330$ &$-27$:$46$:$00.055$ &$0.895$ &$-17.33 \pm 0.11$ &$0.164 \pm 0.003$ &$3.01 \pm 0.07$  \\
$9649$ &$3$:$32$:$40.198$ &$-27$:$46$:$02.859$ &$1.052$ &$-17.21 \pm 0.12$ &$0.426 \pm 0.018$ &$2.64 \pm 0.11$  \\
$9672$ &$3$:$32$:$37.851$ &$-27$:$45$:$22.370$ &$0.925$ &$-17.09 \pm 0.12$ &$0.156 \pm 0.006$ &$3.16 \pm 0.10$  \\
$9712$ &$3$:$32$:$37.887$ &$-27$:$45$:$18.769$ &$0.906$ &$-18.07 \pm 0.08$ &$0.157 \pm 0.002$ &$2.64 \pm 0.06$  \\
$9759$ &$3$:$32$:$38.330$ &$-27$:$45$:$44.208$ &$1.337$ &$-19.23 \pm 0.04$ &$0.644 \pm 0.014$ &$2.06 \pm 0.11$  \\
$9765$ &$3$:$32$:$36.328$ &$-27$:$45$:$41.035$ &$0.978$ &$-17.48 \pm 0.11$ &$0.255 \pm 0.007$ &$2.37 \pm 0.09$  \\
$9783$ &$3$:$32$:$37.349$ &$-27$:$45$:$37.821$ &$1.390$ &$-17.16 \pm 0.12$ &$0.129 \pm 0.008$ &$2.97 \pm 0.11$  \\
$9837$ &$3$:$32$:$36.676$ &$-27$:$45$:$39.144$ &$1.061$ &$-17.22 \pm 0.12$ &$0.279 \pm 0.011$ &$3.04 \pm 0.11$  \\
$9868$ &$3$:$32$:$39.322$ &$-27$:$45$:$32.924$ &$1.150$ &$-17.02 \pm 0.13$ &$0.344 \pm 0.011$ &$2.01 \pm 0.10$  \\
$9962$ &$3$:$32$:$37.496$ &$-27$:$45$:$26.631$ &$0.858$ &$-17.20 \pm 0.12$ &$0.159 \pm 0.003$ &$2.74 \pm 0.07$  \\
$9974$ &$3$:$32$:$38.093$ &$-27$:$45$:$26.839$ &$1.025$ &$-17.62 \pm 0.11$ &$0.504 \pm 0.015$ &$2.50 \pm 0.11$  \\
$9980$ &$3$:$32$:$37.268$ &$-27$:$45$:$28.192$ &$1.170$ &$-17.24 \pm 0.12$ &$0.119 \pm 0.006$ &$2.36 \pm 0.08$  \\
$20037$ &$3$:$32$:$38.245$ &$-27$:$46$:$30.078$ &$1.216$ &$-17.92 \pm 0.09$ &$0.184 \pm 0.004$ &$2.56 \pm 0.08$  \\
$56612$ &$3$:$32$:$36.405$ &$-27$:$45$:$40.645$ &$1.004$ &$-18.92 \pm 0.05$ &$0.122 \pm 0.001$ &$2.39 \pm 0.03$  \\
$56720$ &$3$:$32$:$35.772$ &$-27$:$45$:$48.596$ &$0.699$ &$-17.65 \pm 0.10$ &$0.261 \pm 0.004$ &$2.82 \pm 0.10$  \\
$56815$ &$3$:$32$:$36.197$ &$-27$:$45$:$36.565$ &$0.562$ &$-17.01 \pm 0.13$ &$0.561 \pm 0.005$ &$3.20 \pm 0.10$  \\
$56986$ &$3$:$32$:$30.420$ &$-27$:$46$:$57.275$ &$1.439$ &$-18.80 \pm 0.05$ &$0.162 \pm 0.003$ &$2.68 \pm 0.07$  \\
$57344$ &$3$:$32$:$47.279$ &$-27$:$46$:$50.154$ &$0.678$ &$-17.08 \pm 0.12$ &$0.185 \pm 0.003$ &$2.52 \pm 0.06$  \\
$57390$ &$3$:$32$:$30.949$ &$-27$:$46$:$49.565$ &$1.050$ &$-17.30 \pm 0.11$ &$0.118 \pm 0.008$ &$2.39 \pm 0.11$  \\
$57616$ &$3$:$32$:$34.909$ &$-27$:$46$:$00.389$ &$1.285$ &$-17.99 \pm 0.09$ &$0.253 \pm 0.011$ &$2.98 \pm 0.12$  \\
$58728$ &$3$:$32$:$32.741$ &$-27$:$46$:$30.003$ &$1.307$ &$-18.86 \pm 0.05$ &$0.209 \pm 0.003$ &$2.44 \pm 0.07$  \\
$60189$ &$3$:$32$:$38.153$ &$-27$:$45$:$13.398$ &$0.768$ &$-18.13 \pm 0.08$ &$0.370 \pm 0.004$ &$3.16 \pm 0.08$  \\

\footnotetext[1]{Index in Coe et al. 2006}
\footnotetext[2]{Position based on an optical+IR detection image (Rafelski et al., in prep.)}
\footnotetext[3]{1500~\AA\ rest-frame absolute magnitude}
\footnotetext[4]{Petrosian-like half-light radius at rest-frame 3000~\AA}
\footnotetext[5]{Concentration parameter at rest-frame 3000~\AA}

\end{longtable}

\renewcommand{\thefootnote}{\alph{footnote}}
\newpage

\end{document}